# Coherent Branched Flow in a Two-Dimensional Electron Gas

M.A.Topinka*, B.J.LeRoy†, R.M.Westervelt*†, S.E.J.Shaw†, R.Fleischmann‡, E.J.Heller†§, K.D.Maranowski||, A.C.Gossard||

*Division of Engineering and Applied Sciences, †Department of Physics, § Department of Chemistry and Chemical Biology, Harvard University, Cambridge, Massachusetts 02138, USA  ‡Max-Planck-Institut für Strömungsforschung, Bunsenstraße 10, D-37073 Göttingen, Germany || Materials Department, University of California, Santa Barbara, California 93106, USA

**Semiconductor nanostructures based on two dimensional electron gases (2DEGs) have the potential to provide new approaches to sensing, information processing, and quantum computation.   Much is known about electron transport in 2DEG nanostructures and many remarkable phenomena have been discovered  (e.g. weak localization, quantum chaos, universal conductance fluctuations)[1,2] - yet a fundamental aspect of these devices, namely how electrons move through them, has never been clarified.  Important details about the actual pattern of electron flow are not specified by statistical measures such as the mean free path.  Scanned probe microscope (SPM) measurements allow spatial investigations of nanostructures, and it has recently become possible to directly image electron flow through 2DEG devices using newly developed SPM techniques[3-13].  Here we present SPM images of electron flow from a quantum point contact (QPC) which show unexpected dynamical channeling - the electron flow forms persistent, narrow, branching channels rather than smoothly spreading fans.  Theoretical study of this flow, including electron scattering by impurities and donor atoms, shows that the channels are not due to deep valleys in the potential, but rather are caused by the indirect cumulative effect of small angle scattering.  Surprisingly, the channels are decorated by interference fringes well beyond where the simplest thermal averaging arguments suggest they should be found.  These findings may have important implications for 2DEG physics and for the design of future nanostructure devices.**

Images of electron flow from the quantum point contact (QPC) are obtained by raster scanning a negatively charged SPM tip above the surface of the device and simultaneously measuring the position-dependent conductance of the device. The negatively charged tip capacitively couples to the 2DEG, creating a depletion region that backscatters electron waves. When the tip is positioned over areas with high electron flow from the QPC the conductance is decreased, whereas when the tip is over areas of relatively low electron flow the conductance is unmodified. By raster scanning the tip over the sample and simultaneously recording the effect the tip has on device conductance, a two dimensional image of electron flow can be obtained.

The quantum point contact sample is mounted in an atomic force microscope and cooled to liquid He temperatures. The QPC is formed in the 2DEG inside a GaAs/AlGaAs heterostructure by negatively biasing two gates on the surface – a negative potential on these gates creates two depletion regions that define a variable width channel between them as shown in Fig. 1a. The conductance of the QPC is measured using an ac lock-in amplifier at 11kHz. The heterostructure for the devices used in this experiment was grown by molecular beam epitaxy on an n-type GaAs substrate. The 2DEG resides 57 nm below the surface with mobility $\mu = 1.0 \times 10^6$ cm$^2$/Vs and density n = $4.5 \times 10^{11}$ /cm$^2$. These values of mobility and density correspond to a mean free path $\ell = 11$ μm, Fermi wavelength $\lambda_F = 37$ nm, and Fermi energy $E_F = 16$ meV. The root mean square voltage across the QPC was chosen so as to not heat electrons - 0.2 mV for 1.7K scans. The conductance of the quantum point contact, shown in Fig. 1b, increases as the width of the channel is increased (by changing the gate voltage $V_g$) and shows well defined conductance plateaus at integer multiples of the conductance quantum $2e^2/h$[1,2]. When probing the electron flow, the SPM tip was held at a negative potential relative to the 2DEG and was scanned at 10nm above the surface of the heterostructure.

Figures 2a and 2b show images of electron flow from two different quantum point contacts at the temperature 1.7K; both QPCs are biased on the G = $2e^2/h$ conductance plateau. Figure 2b shows the flow patterns on each side of a quantum point contact (the gated region at the center was not scanned), and Figure 2a shows a higher-resolution image of flow from one side of a different QPC. In both these images, the current exits the point contact in a central lobe, as would be expected from an exact quantum-mechanical calculation of electron flow through an ideal QPC without impurities or non-uniform distributions of dopant atoms. Rather than continuing out as a smoothly widening fan, it quickly forks into several different paths and continues to branch off into ever smaller rivulets for the full width of the scan. This branching behavior was observed in all of the 13 QPC exit patterns observed so far. Previously, there have been suggestions of an unexpected narrowness in observed flow from a QPC[6], but until now high-resolution, detailed images of electron flow from a quantum point contact have been difficult to obtain and no observations or predictions for this type of strong branching behavior have been published. It is interesting to note that all flow images (Fig 2a and 2b) are independent of current direction. The average electron flow reflected by the tip falls off approximately as $1/r^2$ with distance r from the QPC.

In order to explore the experimentally observed channeling and branching of current flow, we numerically calculated a set of conducting wavefunctions through a quantum point contact with disordered background potentials. Our model for the potential landscape incorporates known properties of 2DEGs in GaAs/AlGaAs heterostructures[14,15] and replicates the physical parameters of the experimental system. To model the full in-plane potential, treating the 2DEG as an idealized δ-layer, we consider two contributions. The first contribution is from the negatively charged gates which define the QPC, which we model by a smooth analytic potential that reproduces the quantized conductance steps characteristic of the QPC. The Fermi level remains constant inside the 2DEG, so that the propagation there is primarily affected by the second piece of the potential, the disordered background. There are two contributions to this background that we considered: the donor dopant atoms and unwanted impurities. For the impurities, distributed throughout the crystal structure, we take a random distribution in three dimensions and match the impurity atom density with a reasonable estimate for the physical sample ($1.25 \times 10^{15}$ /cm$^3$). The donors are located in a δ-layer displaced from the 2DEG by 22nm with a sheet density of $8 \times 10^{12}$/cm$^2$; we estimate that half of the donors are ionized. The distribution is random in space except for correlations limiting the maximum local density[14]. We use a $1/r^3$ potential at large distances for the impurities and donors, where r is the distance between a location in the plane of the 2DEG and the location of a given impurity or donor atom in the crystal. This $1/r^3$ dependence is the key feature of the full screened potential in a 2DEG from a point charge[15].

Using the values for donor inhomogeneity and impurity density expected for our 2DEG sample in our model[14], we calculated the mobility classically via the momentum relaxation time for an appropriate ensemble of trajectories and found it to match the measured value to within 10%. The distribution of energies in the disordered potential is very nearly Gaussian with a standard deviation of about 8% $E_F$. The correlation length of the disordered potential is about 25 nm, i.e. on the order of the wavelength.

In Fig. 3a we show a typical potential, including the portion from the point contact gates, used in both classical and quantum simulations. Figures 3b and 3c show the results of classical and quantum-mechanical calculations in that potential. In the classical results, we followed the dynamics of an appropriate ensemble of classical trajectories and show the classical flux density. The quantum-mechanical results show the flux density of the transmitted wavefunction, coming through the point contact on the left and going into free space on the right. The agreement between the classical and quantum results is very good, leading to the conclusion that the branched flow is essentially a classical phenomenon. The "shadow" in Fig. 3a, cast by the classical flux simulations from Fig. 3b, demonstrates that channeling is not simply caused by large valleys in the potential - the current paths do not simply follow low potential routes which guide the current as a valley would guide a river. Though there are occasional events where the flow is split by an impurity near the 2DEG, most of the bumps in the potential which cause channeling are well below the Fermi energy of the electrons, so the flow does not represent the only energetically available paths. Rather, they are produced by a folding of electron trajectories in phase space resulting in cusp caustics in coordinate space, which are caused by the focussing effect of potential dips. One example of such a cusp in the experimental images may be seen in Fig. 2a as indicated by the arrow. The stronger branches surviving at large distances are the result of small-angle scattering and trajectories of marginal instability, with stable axes directed along the initial distribution of trajectories in phase space. These longer, occasionally branching filaments are dependent on all the fine structure in the potential "upstream" of a given point in the filament.

This classical phenomenon is robust, and it is seen in smooth potentials of disparate origin with features generally well below the energy of the scattered particles. The length scale for the formation of the channels is determined by the autocorrelation length of the potential. This regime of classical dynamics in a random potential has not been studied in detail, falling as it does between the well-explored extremes of Gaussian white noise and widely spaced point scatterers. A detailed study of this phenomenon will appear elsewhere[16]. Similar scattering, with the unexpected resulting strands of concentration of flux, might be expected in widely different contexts, such as sound propagation through the ocean, where it has recently been discussed[17].

The experimental images in Fig. 2 show the position-dependent effect of the SPM tip on the conductance through the device. The charged tip induces an approximately Lorentzian bump in the potential[18] seen by electrons propagating through the system; we can thus simulate the experiments computationally by introducing an appropriate Lorentzian into the potential and finding the conductance as a function of its position. Figure 4a shows the overall computed current flow through the device before the addition of a Lorentzian. We compared the simulated flux in a small patch (Fig. 4b) with the conductance as a function of Lorentzian position (Fig. 4c). This comparison confirms the relationship between flux and the images achieved by the experimental technique.

A striking feature of the experimental images (Figs.2a and 2b) is the appearance of fringes oriented perpendicularly to electron flow and spaced by $\lambda_F/2$, half the Fermi wavelength. These fringes are due to coherent constructive and destructive backscattering of electrons from the tip. In the region close to the point contact the fringes can be explained by multiple reflections of electron waves between the tip and the QPC gates. At farther distances there are two effects that could lead to the suppression of the fringes; the loss of phase coherence and the dispersion in wavelength caused by thermal broadening. In our sample the phase coherence length $\ell_\phi = 18\mu m$[19] which is much larger than our scan area. However the thermal length $\ell_{th} = \hbar^2 2\pi/m\lambda k_B T = 1.4\mu m$. One might expect the relative strength of the ripples to die out exponentially with a decay length equal to $\ell_{th}/2$. For our system this would correspond to a factor of 1/e decrease in relative fringe amplitude at a distance of 0.7μm from the QPC. The fringes in our experimental images, however, persist up to this length and well beyond.

The surprising persistence of the fringes past the thermal length can be explained by an interesting and unusual source of coherent backscattering. The electron waves backscattered from the depletion region under the tip, (located a distance $r_{tip}$ from the QPC), interfere with a combination of the backscattered waves from all impurities in a ring covering the area $r_{tip} \pm \ell_{th}$ away from the QPC to produce alternating constructive and destructive total interference as $r_{tip}$ is changed[20]. (Scatterers outside this ring do not contribute as strongly, because of thermal averaging.) As the tip is moved, the phase of the waves backscattered from the tip seen at the QPC varies as $2 r_{tip} k$, where k is the wavevector and $r_{tip}$ is the distance from the QPC to the tip. Importantly, this mechanism is resistant to thermal broadening, because as k is varied around $k_F$ the phases of backscattered waves returning from impurities within the radius $r_{tip} \pm \ell_{th}$ vary in step with the phase from the tip. The result is that the fringes can easily survive farther away from the QPC than the 1.7K thermal broadening length. Simple numerical simulations involving point scatterers bear out this model. This mechanism requires phase coherent transport over the round-trip distance to the most remote fringes, several microns. These fringes not only demonstrate that it is possible for coherent backscattering to be resistant to thermal smearing, but may also provide a new direct way to measure electron wave coherence length.

A second surprise seen in the fringes is that their separations are smaller than the width of the Lorentzian perturbation used in making the measurements. The tip perturbation has an estimated half-width at half-maximum of ~60 nm, based on electrostatic simulations[18,21]. Our numerical work, however, gives evidence that the relevant feature is the presence of a classically forbidden "depletion region" beneath the tip[16]. The edge of this depletion region, rather than the tails of the potential, provides the backscattering needed to direct electrons back through the QPC. This result is in accord with experimental findings that images of electron flow are only observed when the voltage on the tip is sufficient to create a small depletion region in the 2DEG[21].

The information about electron flow made available by this imaging technique has a number of important implications and opens possible future applications. The observation that electrons can flow through a 2DEG in narrow, branching channels can be important for experiments relying on the assumption of "ballistic" flow of electrons for distances less than the mean-free-path. While the ballistic assumption is an excellent starting point, it may also be important to include the unexpected channeling of flow reported here. The detailed images of electron flow available from this imaging technique may also prove valuable in future explorations of spin transport for spintronics and possibly allow novel quantum computer implementations. The surprising persistence of coherent fringes well past the thermal length may help provide additional insight into a broad range of coherent phenomena including universal conductance fluctuations, phase coherence, and weak localization. This imaging technique is capable of providing a high level of detail in imaging coherent electron flow, and is an important tool in investigating the underlying physics as well as the future design of 2DEG nanostructures.

Supported in part at Harvard by Office of Naval Research/Augmentation Awards for Science and Engineering Research Training (ONR/AASERT), by ONR and the National Science Foundation through grants for the Materials Research Science and Engineering Center and for the Institute for Theoretical Atomic and Molecular Physics at Harvard University and Smithsonian Astrophysical Observatory. Work at the University of California Santa Barbara was supported by NSF Science and Technology Center QUEST.

**Correspondence should be addressed to R.M.W. (e-mail: westervelt@deas.harvard.edu).**

**Figure 1** Experimental setup **a** Schematic diagram of the experimental setup used for imaging electron flow. The tip introduces a moveable depletion region which scatters electron waves flowing from the quantum point contact (QPC). An image of electron flow is obtained by measuring the effect the tip has on QPC conductance as a function of tip position. **b** Conductance of the QPC used for Fig. 2b vs. QPC width controlled by the gate voltage. Steps at integer multiples of $2e^2/h$ are clearly visible.

**Figure 2** Experimental images of electron flow **a** Image of electron flow from one side of a QPC at T = 1.7K, biased on the $G = 2e^2/h$ conductance step. Dark regions correspond to areas where the tip had little effect on QPC conductance, and hence are areas of low electron flow. The color varies and the height in the scan increases with increasing electron flow. Narrow branching channels of electron flow are visible, and fringes spaced by $\lambda_F/2$, half the Fermi wavelength, are seen to persist across the entire scan. **b** Images of electron flow from both sides of a different QPC, again biased on the $G = 2e^2/h$ conductance step. Strong channeling and branching are again clearly visible. The white arrow points out one example of the formation of a cusp downstream from a dip in the potential.

**Figure 3** Calculated electron flow **a** Surface plot of the random potential for computed electron flow, including contributions from impurities and donors; green areas are low and white areas are high potential. The "shadow" is cast by classical flux through the same potential. Note that the branched flux does not follow channels in the potential. **b** Classical and **c** quantum-mechanical flux of electrons flowing through the potential in **a**. Note that both results show the same branching behavior.

**Figure 4** Calculated tip scan **a** Quantum-mechanical flux through a random potential. **b** The flux from the boxed area in **a**. **c** A raster scan of conductance as a function of SPM tip position in the same system as **a** and **b**. The conductance image in the model corresponds to the flux image, confirming our assertion that the experiment images electron flow. Additionally, the simulationin **c** shows quantum fringes, as seen in the experiment. Though this simulation is at zero temperature, the fringes do survive thermal averaging.

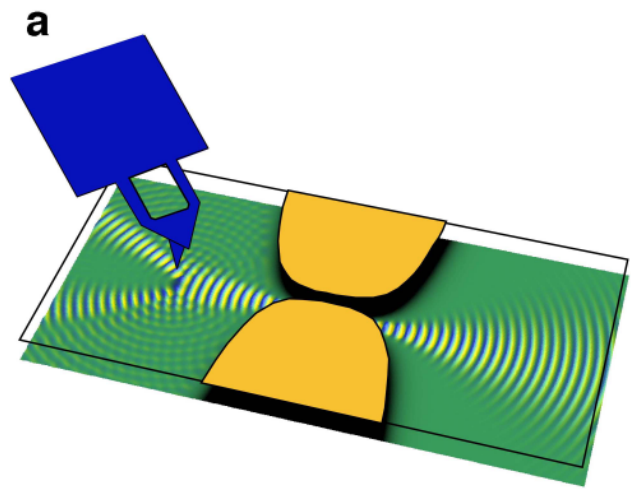

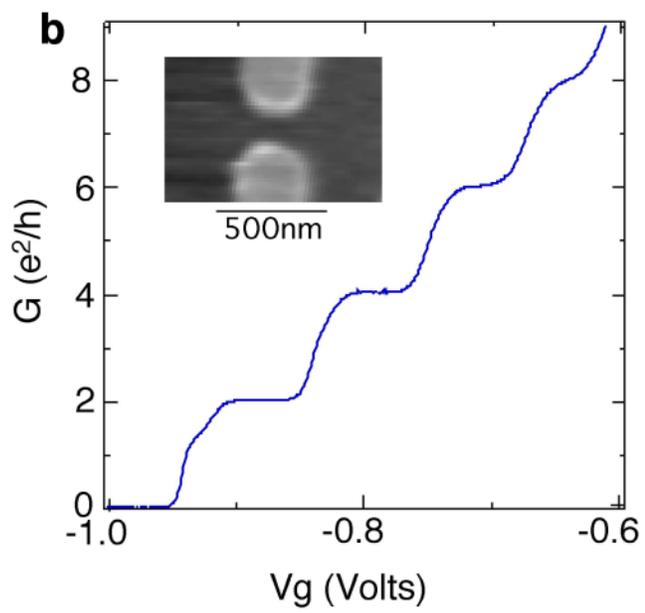

Figure 1
Topinka, M.A.

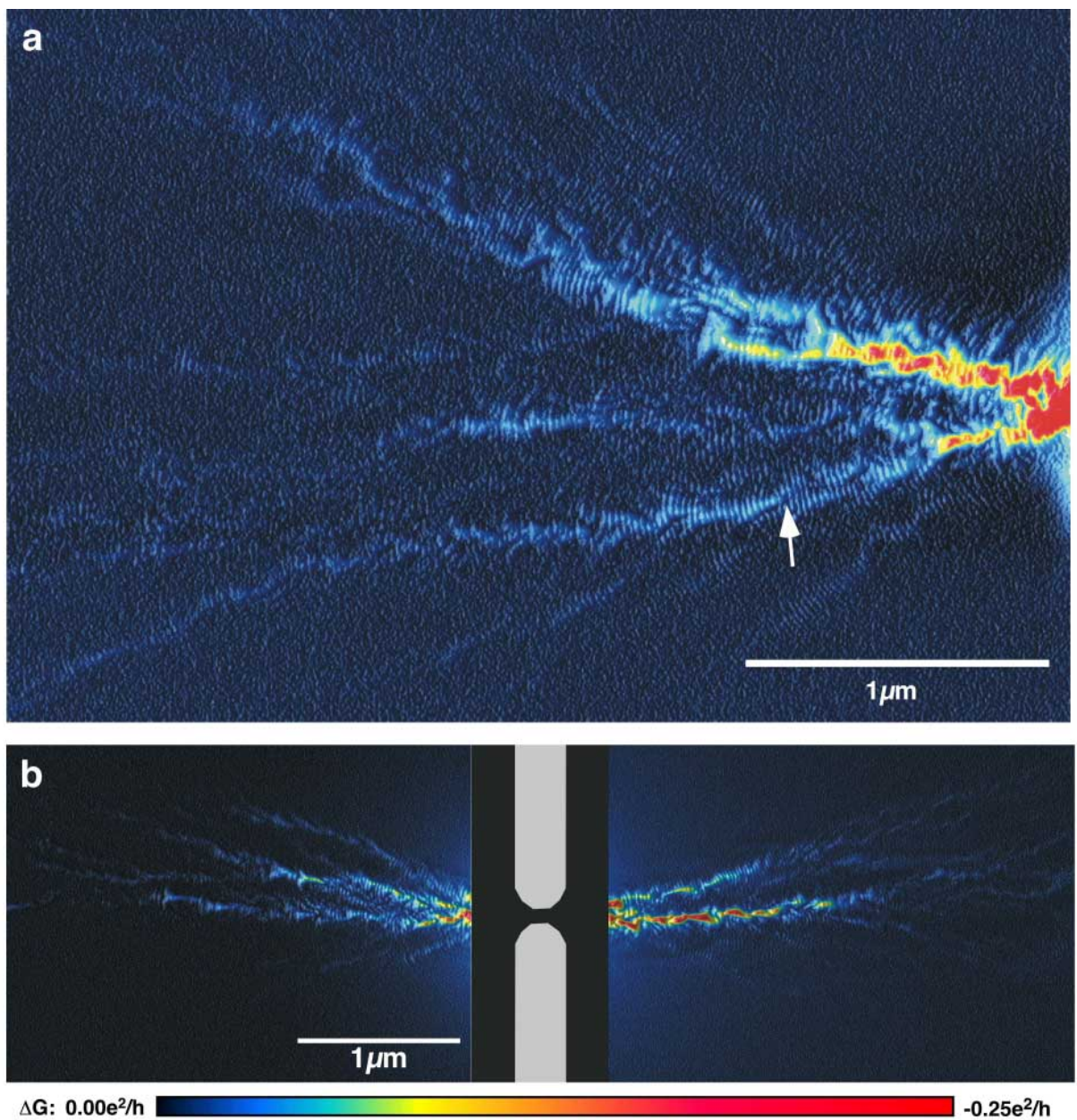

Figure 2
Topinka, M.A.

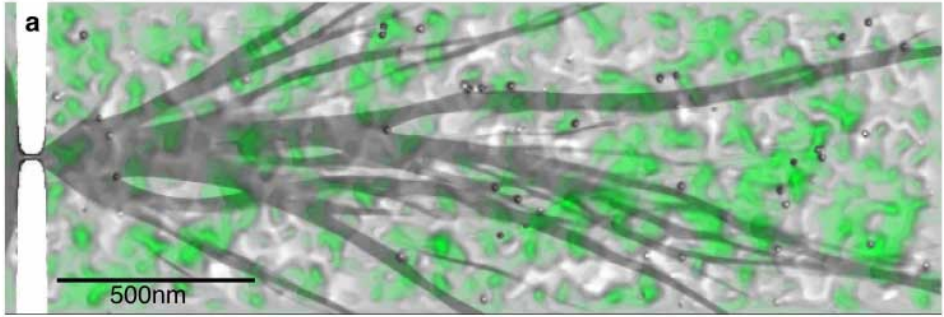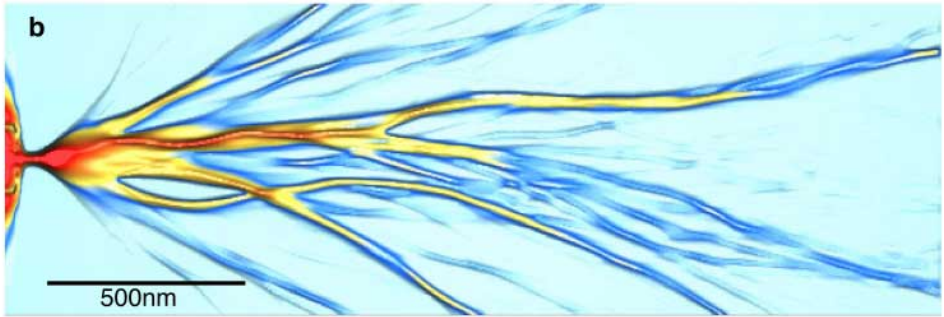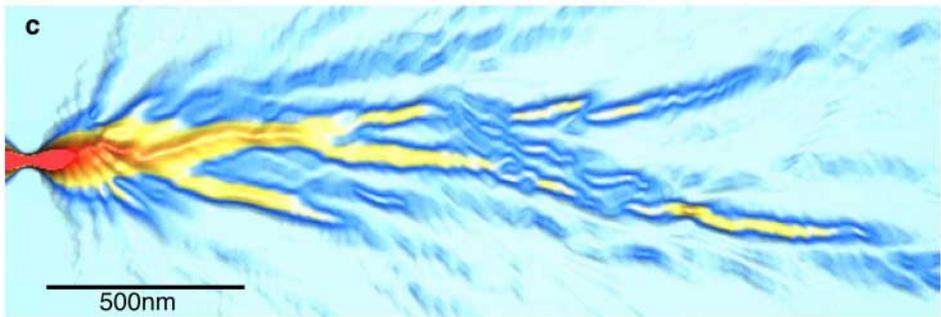

Figure 3
Topinka, M.A.

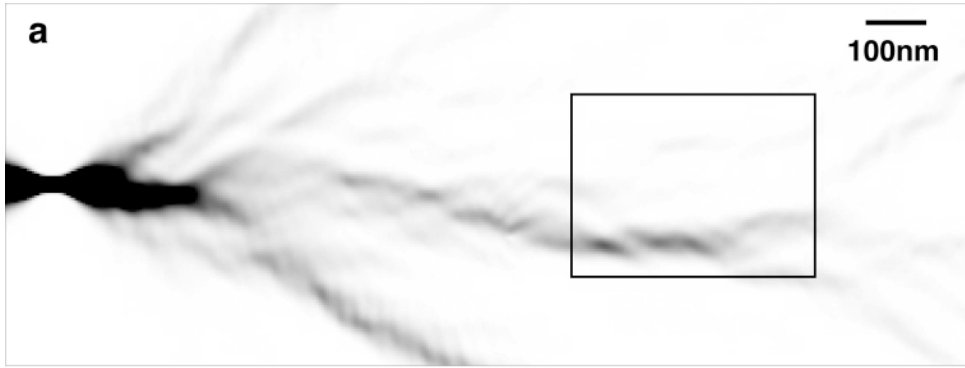
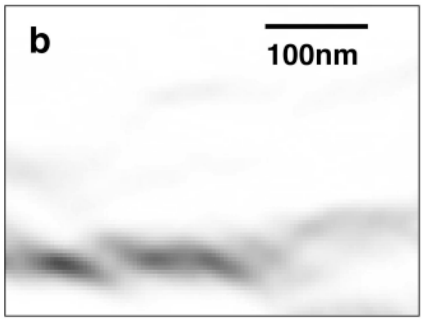 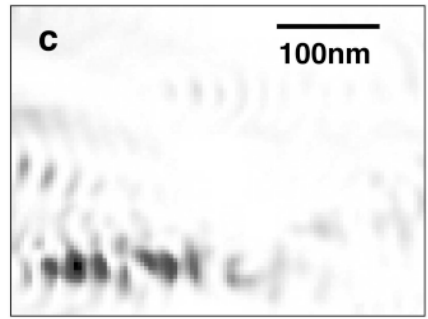

Figure 4
Topinka, M.A.